\documentclass[preprint,12pt]{elsarticle}

\usepackage{dcolumn}% Align table columns on decimal point
\usepackage{bm}% bold math
\usepackage[usenames]{color} % colors

\begin{document}

\title{Pseudopotentials for high-throughput DFT calculations}% Force line breaks with \\

\author[rutgers]{Kevin F. Garrity}
\ead{kgarrity@physics.rutgers.edu}
\author[rutgers]{Joseph W. Bennett}
\author[rutgers]{Karin M. Rabe}
\author[rutgers]{David Vanderbilt}
\address[rutgers]{%
Department of Physics \& Astronomy, Rutgers University, Piscataway, New Jersey 08854-8019, USA\\
}%

\date{\today}% It is always \today, today,

\begin{abstract}
The increasing use of high-throughput density-functional theory
(DFT) calculations in the computational design and optimization
of materials requires the availability of a comprehensive set of
soft and transferable pseudopotentials.  Here we present design
criteria and testing results for a new open-source ``GBRV''
ultrasoft pseudopotential library that has been optimized for
use in high-throughput DFT calculations.  We benchmark the GBRV
potentials, as well as two other pseudopotential sets available in
the literature, to all-electron calculations in order to validate
their accuracy.  The results allow us to draw conclusions about
the accuracy of modern pseudopotentials in a variety of chemical
environments.
\end{abstract}

%\pacs{68.47.Fg, 71.15.Mb, 81.15.Hi}% PACS, the Physics and Astronomy
\maketitle

\section{\label{sec:intro}Introduction}

The use of pseudopotentials for practical and efficient
electronic-structure calculations has a long history in computational
condensed-matter physics.\cite{psp1,psp2,psp3,psp4,psp5} In
pseudopotential-based electronic structure calculations, the nuclear
potential and core electrons of an atom are replaced by a much softer
effective potential felt by the valence electrons, enabling the latter
to be described with a relatively small number of basis functions.
Thus, the success and popularity of various first-principles
density-functional theory\cite{hk,ks} (DFT) codes using plane-wave
basis sets (e.g. VASP,\cite{vasp, vasp2} QUANTUM-ESPRESSO,\cite{QE}
ABINIT\cite{abinit}) depend on the availability of high-quality
pseudopotentials.  While the publication and testing of entire pseudopotential
libraries also has a long history,\cite{pspthatwork, tm_set, hgh,
joe_psp, delta} the dominant mode of pseudopotential design and testing has
been that of case-by-case construction, with authors typically
creating and testing potentials only for a specific application.  This
\textit{ad hoc} method of pseudopotential design and testing is
incompatible with the increasing use of first-principles DFT
calculations in materials design applications, especially those that
make use of high-throughput calculations to explore the properties of
(possibly hypothetical) materials constructed from atoms across the
periodic table.\cite{wolverton, ceder, ceder2, curtarolo, curtarolo2,
evolutionary, cluster, datamine, thermoelectric, halfheusler,
abcferro, abcantiferro, zunger_abc} Moreover, the pseudopotentials
themselves (or the input data needed to construct them) are often not
posted or published, and it is even more rare for comprehensive
pseudopotential testing data to be made publicly available.
Unfortunately, this state of affairs creates practical difficulties in
the duplication of previous studies and limits our understanding of
the accuracy and transferability of modern pseudopotentials in
realistic calculations.  In addition, the lack of an open-source
pseudopotential library appropriate for high-quality high-throughput
calculations likely limits the adoption of this technique.

In this work we introduce a new ``GBRV'' open-source pseudopotential
library, explaining the design criteria used to construct it and
providing a suite of test results to verify its accuracy.  The GBRV
library, available at http://physics.rutgers.edu/gbrv, consists of
highly accurate ultrasoft\cite{ultrasoft} pseudopotentials generated
using the Vanderbilt pseudopotential generation
code.\cite{ultrasoft_code} We provide input files for the
pseudopotential generator as well as ultrasoft pseudopotential files
which can be used directly with QUANTUM ESPRESSO and
projector-augmented wave (PAW)\cite{paw} files generated with the
uspp2abinit\cite{uspp2abinit} add-on which can be used directly
with ABINIT.  Our library has been designed and optimized for use in
high-throughput DFT calculations, though it should be appropriate for
many applications.  In addition, we test two other PAW libraries, the
mature but proprietary VASP PAW library version 5.2\cite{paw2} and the
still-under-development PSLIB 0.3.0 public PAW library\cite{pslib,
pslib_poster} generated using the ATOMIC code of the QUANTUM ESPRESSO
package.  Versions of the VASP library have been used in the majority
of previous pseudopotential-based high-throughput studies, usually
with little discussion of its accuracy.\cite{ceder, ceder2, curtarolo,
curtarolo2, wolverton, datamine, cluster, thermoelectric, zunger_abc}

Testing three potential sets allows us to assess the accuracy of
our library relative to other choices and also provides some perspective
on the limits
of current pseudopotential methodology.  We have tested the potentials by
comparing with all-electron (AE) results from WIEN2k,\cite{wien2k}
which uses the highly accurate full-potential linearized augmented
plane-wave + local orbitals method (FLAPW+LO).

The manuscript is organized as follows.  Our pseudopotential design
criteria are presented and discussed in Sec.~II.  The tests of
the GBRV, VASP and PSLIB potentials in comparison with AE calculations
are presented in Sec.~III.  Our conclusions are summarized in Sec.~IV.

\section{\label{sec:design} Pseudopotential Design}

\subsection{Design criteria for high-throughput}

High-throughput DFT studies of materials systems present a variety of
challenges for pseudopotential design which have informed our choices in
creating the GBRV pseudopotential library.  First, by their very nature,
high-throughput calculations include a wide variety of atoms and
thus require accurate potentials extending throughout the periodic
table.  Therefore, our library includes all of the
elements from H to Bi except for the noble gases and the $f$-block
elements.

Second, high-throughput calculations require significant computational
resources, which should be minimized if possible.  Therefore, we
designed our potential library so that all of the elements can be run at
a relatively low plane-wave cutoff of 40 Ry and a charge-density
cutoff of 200 Ry.  Using a single low cutoff greatly simplifies
the use of our potentials in high-throughput calculations.  This
contrasts with the normal procedure of allowing a variety of
plane-wave cutoffs in a single pseudopotential library.
In that case, one either has to use the highest cutoff in
all calculations or else face difficulty in comparing energies of
structures containing different combinations of atoms, a procedure
that is central to the kind of thermodynamic stability analysis
that is often required in high-throughput calculations.

Third, high-throughput calculations often place atoms in unusual or
novel chemical environments.  Therefore, we required that our
potentials be highly transferable and able to reproduce metallic,
ionic, and covalent bonding behavior.  For the purposes of reliable
high-throughput calculations, it is crucial that the potential library
produces consistently accurate results for all atoms in all reasonable
crystal structures.  This requirement led us to include extra
semi-core states in normally borderline cases, as a high-throughput
study cannot verify whether the semicore states are necessary in every
structure examined.

Finally, on a more technical level,
high-throughput calculations of bulk materials typically
require the use of variable unit-cell relaxation algorithms to
minimize the stress and optimize lattice parameters.  These
calculations present numerical difficulties as they are normally run
at a fixed number of plane-waves determined by the initial
configuration, rather than at a fixed plane-wave cutoff, and they
require accurate determination of stresses, which can be expensive to
converge.  Therefore, we designed our potentials such that when
used with a smeared plane-wave cutoff they both produce accurate
stresses and converge to an accurate ground state structure when
using a variable cell relaxation algorithm (given a reasonable
starting structure).

\subsection{Pseudopotential construction}

Unfortunately, designing a set of pseudopotentials which meets all of
the above requirements is very difficult, as the requirements are
naturally in conflict with each other.  Highly transferable potentials
generally require high plane-wave cutoffs and many semi-core states,
which is in direct conflict with the requirement of a low plane-wave
cutoff.  In addition, a comprehensive set of potentials is both more
difficult to design with a single low plane-wave cutoff and allows for
many chemical environments, making reliability difficult.  Given these
conflicts, one is naturally led to adopt either ultrasoft or PAW
potentials, which can provide both higher transferability and lower
plane-wave cutoffs than norm-conserving potentials.  In the present
work, we have chosen to design a library of ultrasoft
pseudopotentials.  We describe our procedure for optimizing potentials
below; however, there remain atoms which are particularly difficult to
pseudize given the above constraints, which we discuss further when we
present our testing data in section \ref{sec:results}.

The process of construction of the potentials consisted of optimizing the
following parameters: (a) a reference atomic configuration (neutral or
slightly ionic), (b) the number of core and valence states, (c) the
cutoff radii for each angular momentum channel, (d) a local potential
and local cutoff radii, (e) the inner pseudization and non-linear core
correction radii, and (f) the energies of any extra
projectors.\cite{psp_gen}  We began our design by constructing initial
potentials from previously-tested potentials if available, using
periodic trends to fill in missing elements, and testing each atom first in
\textit{fcc} and \textit{bcc} structures and then using the NaCl testing set 
(see section \ref{sec:details}).  We found that expanding our testing to 
the perovskites and
half-Heuslers required relatively little additional tuning.  While the
traditional transferability tests provided by pseudopotential
generators, such as comparing the logarithmic derivatives to
all-electron results and testing the pseudopotential in multiple
atomic configurations, were helpful in narrowing the choices of
parameters for our potentials, we found that these tests are rarely
sufficient to design a potential which meets the competing goals of
our design criteria.

The various parameters of our potentials were adjusted by hand with
extensive use of solid-state testing to identify which atoms need
improvement and which aspects of those potentials must be modified.
When a potential performed poorly in tests, we first adjusted the outer cutoff
radii by deciding if the potential was either too hard or too soft.  A
potential which is too hard will improve its performance when tested
with higher plane-wave and charge-density cutoffs, and requires
increasing the cutoff radii of the local potential or non-local
projectors, while an excessively soft potential must be adjusted in
the opposite direction.  Cutoff radii tend to follow periodic trends,
with radii decreasing as one moves right across a row and increasing
as one moves down a column, although differences in numbers of
semicore states complicates this relation.

If an element had poor testing results which were found to be
insensitive to the projector cutoff radii, there were several other
options which we considered.  First, we added any additional relevant
semicore states or a non-linear core correction if there was
significant valence/core overlap.  Second, we looked in detail at the
logarithmic derivatives and adjusted the local potential and the
energies of any extra projectors in order to improve agreement in the
chemically relevant region.  Almost all of our occupied angular
momentum channels have at least two projectors, and a few with
semi-core states have three.  We found that potentials heavier than
fluorine generally require a good description of the $d$ channel and
the bottom 1-2 rows of the periodic table require a reasonable description
of the $f$ channel, especially for atoms near La-Hf.  Third, it was sometimes 
necessary to adjust the inner pseudoization radius; magnetic moments
are often particularly sensitive to this parameter.  For most
elements, we were able to achieve potentials which met all of our
requirements after a few revisions, and small changes in the parameters
would not affect the testing results noticeably.  For a few
problematic elements (Li, Be, F, Na, Ni, Hf) we were forced to
systematically vary and test certain parameters that were found to be
particularly important in balancing the trade-off between
accuracy and hardness.  In some cases we also had
to adjust the local potential or
reference configuration in order to find a region of parameter space
which met our requirements as closely as possible.  Having a large
solid state testing set was important to prevent over-fitting of these
difficult cases to any particular property.  We discuss some of these
atoms further in section \ref{sec:results}.

\section{\label{sec:testing} Tests}

% % \subsection{Testing set details}
% \dvm{Previous subsection title was ``Testing set details'' but this doesn't
% come until the last paragraph.  I think we can start the section
% without a subsection.}

In order to gauge the transferability of a general-purpose
pseudopotential library, it is necessary to test the potentials in
chemical environments that include ionic, covalent, and metallic
bonding.  For this purpose, we have chosen a testing regimen
in which the pseudopotential calculations are compared not with
experiment, but with all-electron calculations performed under
conditions that are as identical as possible.  This allows us to
quantify the accuracy of the pseudopotentials themselves, isolated
from any complications having to do with actual experimental
conditions (finite temperature, zero-point motion, etc.)
or with theoretical approximations that are common to both the
pseudopotential and all-electron calculations.

Thus, our pseudopotential and all-electron calculations
are always carried out with exactly the same choice of DFT
exchange-correlation functional, the same k-point sets, and the same smearing temperature.
The PBE exchange-correlation functional\cite{pbe} was used throughout.
Both the AE calculations and the pseudopotential constructions were
scalar-relativistic,\cite{kh}
i.e., without spin-orbit interactions.\footnote{WIEN2K treats
core electrons fully-relativistically, while the pseudopotentials treat
the core electrons scalar-relativistically\cite{kh},
which may result in small systematic errors.}
We ran all of our testing calculations as non-spin-polarized
calculations, except for our calculation of the magnetic moments
of the transition metal oxides, which we ran at the all-electron
non-spin-polarized lattice constants.

In the same spirit, we can reduce the computational load associated
with the test suite by making some common approximations that still
allow systematic comparison.
For example,
we ran all of our calculations on an 8$\times$8$\times$8 $k$-point
density and with 0.002 Ry Fermi-Dirac smearing.  We note that this
$k$-point and temperature setting is not sufficient to fully converge
many of the properties we calculate and, as stated above, the results
should not be compared with experiment. However, by using the same
setting for all calculations, we can nevertheless compare the results
to each other on an equal footing.

\subsection{Testing procedure}

Calculations with the GBRV
pseudopotential set were run using QUANTUM ESPRESSO at our target
plane-wave cutoff of 40 Ry and a charge density cutoff of 200 Ry.
The PSLIB set provides a variety of suggested plane-wave cutoffs which
range up to 78 Ry, but most are below 50 Ry and we ran all
calculations with a cutoff of 50 Ry and a charge density cutoff of
450 Ry, also using QUANTUM ESPRESSO.  This lower-than-recommended
cutoff may bias results against PSLIB, but we think a relatively
low cutoff is appropriate given our goal of testing potentials for
high-throughput applications.

VASP provides a variety of potentials
for each atom; we chose the potentials with the largest number of
semi-core states (we did not test the new `GW' potentials).  These
potentials have cutoffs of up to 47 Ry, although most are below 40
Ry.  For cases such as O and F which include soft, normal, and hard
versions, we chose the normal version.  We ran all VASP calculations
using the `high' precision setting, which increases the plane-wave
cutoff 25\% above the suggested values, which is necessary to converge
the stress for variable cell calculations.

WIEN2K calculations were
performed at $R_{\rm MT}K_{\rm MAX}$ values of 8-10 as necessary to converge the lattice constant.  Calculations of lattice
constants and bulk moduli proceeded by first performing a variable-cell
relaxation with the GBRV pseudopotential set to generate an initial
guess for the lattice constant, and then performing energy
calculations with each of our testing sets at nine points from $-$1\% to
1\% of the initial guess and fitting the results to a parabola.

\subsection{\label{sec:details}Details of the testing sets}

Our four testing sets all consist of cubic materials without free
internal parameters in order to reduce the computational demands
associated with structural relaxation.  Our first testing set consists
of each of the elements in simple \textit{fcc} and \textit{bcc}
structures.  This set was designed to test the potentials in a
metallic bonding environment as well as to allow us to analyze each
potential separately, although we note that for many elements these
structures are highly unrealistic and represent very demanding test
cases.\footnote{We were unable to converge WIEN2K calculations of the
\textit{fcc} and \textit{bcc} structures of N, P, or Hg.}  The second
testing set consists of rocksalt structures designed to test ionic
bonding.  Main group elements are paired to put both elements in
preferred oxidation states; most of these structures are insulating.
All of the transition metal elements, which often have multiple
oxidation states, are paired with oxygen; many of these are metallic.
The third testing set consists of members of the heavily-investigated
perovskite family.  This set also largely tests ionic bonding, but
includes tests of elements in higher oxidation states than the
rocksalt structures.  Finally, we test a large set of half-Heusler
structures (MgAgAs structure type, space group
$F\bar{4}3m$).\cite{simplehh,halfheusler} Half-Heuslers display a
complicated combination of ionic, covalent, and metallic bonding, and
should give an indication of the accuracy of our potentials in a
variety of realistic environments.  Both half-Heuslers and perovskites
were chosen for their simple structure and their common inclusion of
elements from throughout the periodic table.  We include almost ninety
previously synthesized half-Heuslers plus additional hypothetical
examples, bringing our half-Heusler test set to 138 compounds, which
include all of the elements in our pseudopotential set except H and
the halogens.

\subsection{\label{sec:results}Results}

\begin{table} 
\caption{\label{tab:summary} Summary of pseudopotential testing results.  All
testing data is presented as either RMS errors relative to AE
calculations or percent of lattice constants outside $\pm 0.2 \%$ .  Only compounds where all
three pseudopotential sets converge are included in RMS errors.}
%\begin{ruledtabular}
\begin{tabular}{llll}
Test  & GBRV  & VASP & PSLIB   \\
\hline
\textit{fcc} latt. const. (\%) & 0.14  & 0.13  & 0.36  \\
\textit{bcc} latt. const. (\%) & 0.15  & 0.13  & 0.28  \\
\textit{fcc} bulk modulus (\%) & 3.6  & 4.1  & 5.4  \\
\textit{bcc} bulk modulus (\%) & 5.3  & 4.6  & 7.1  \\
\textit{fcc}-\textit{bcc} Energy (meV) & 3.7  & 3.6  & 5.4  \\ 
rock salt latt. const. (\%) & 0.13 & 0.15 & 0.21 \\
rock salt bulk modulus (\%) & 5.0 & 4.5 & 4.9 \\
rock salt mag. mom. ($\mu_{B}$) & 0.08 & 0.22 & 0.01 \\
perovskite latt. const. (\%) & 0.08 & 0.13 & 0.20 \\
perovskite bulk modulus (\%) & 5.5 & 6.1 & 7.7 \\
half-Heusler latt. const. (\%) & 0.11  & 0.14  & 0.13 \\
half-Heusler  bulk modulus (\%) & 5.4 & 5.8 & 5.1 \\
\hline
\textit{fcc} latt. const. $>\pm 0.2 \%$ (\%) & 9.8  & 9.8  & 26.2  \\
\textit{bcc} latt. const. $>\pm 0.2 \%$ (\%) & 9.8  & 8.2  & 27.9  \\
rock salt latt. const. $>\pm 0.2 \%$ (\%) & 7.8 & 14.3 & 26.6 \\
perovskite latt. const. $>\pm 0.2 \%$ (\%) & 0.0 & 14.5 & 27.3 \\
half-Heusler latt. const. $>\pm 0.2 \%$ (\%) & 3.6 & 15.2  & 13.6 \\
\end{tabular}
%\end{ruledtabular}
\end{table}

In Table \ref{tab:summary},
we present summary data for the overall performance the three
pseudopotential sets (see supplementary materials for more details).  Each line in the table summarizes either the
root-mean-squared (RMS) error relative to the AE calculations for a
given type of test (excluding any cases in which there was a
convergence failure for any of the pseudopotentials), or the
percentage of structures in the testing set which have lattice
constants errors outside the range of $\pm 0.2\%$.
Given our goal of transferability and reliability across a wide
variety of structures as well as across the periodic table, this last
measure is important to assess the robustness of the pseudopotential
sets, rather than focusing only on averaged results, as reliable 
results require every calculation to be accurate.  In general, we find that
all three sets perform well, with most lattice constants within 0.2\%
of the all-electron results, and most bulk moduli within 5\%.  In
fact, for many materials the differences between the pseudopotentials
and all-electron calculations are comparable to uncertainties in the
all-electron results themselves.  For the properties included in our
testing set, the aggregate performance of the GBRV pseudopotential set
is superior to the VASP PAW library,
especially with regards to the robustness of results for structures
containing multiple elements in covalent and ionic environments.  Both
the GBRV and VASP sets give better lattice constants than PSLIB, but
the bulk modulii and energy differences are similar for all three testing sets.  PSLIB gives highly accurate
magnetic moments, while VASP has a few elements which give poor magnetic moments.

The results for all three sets are clearly superior to those for
norm-conserving pseudopotential libraries such as the TM\cite{tm_set}
and HGH\cite{hgh} sets, despite having significantly softer
plane-wave cutoffs.  For example, Fig.~5 of Ref.~\cite{joe_psp}
compares the performance of some of these older comprehensive
norm-conserving libraries against a recently-developed
``Bennett-Rappe'' library, using a test of half-Heusler lattice
constants similar to that shown in the penultimate line of
Table \ref{tab:summary}.  The RMS lattice constant errors of 2.3\%
and 2.8\%
for the TM and HGH sets respectively were reduced to
only 0.76\% for the Bennett-Rappe set, with some of the most
significant improvements occurring for transition-metal elements;
this probably approaches the
limit of what can be achieved in a single-projector norm-conserving
framework.  The results for the ultrasoft and PAW libraries
in Table \ref{tab:summary}, however, provide a dramatic additional
reduction to less than  0.15\% RMS error.

\begin{figure*}
\includegraphics[width=5in]{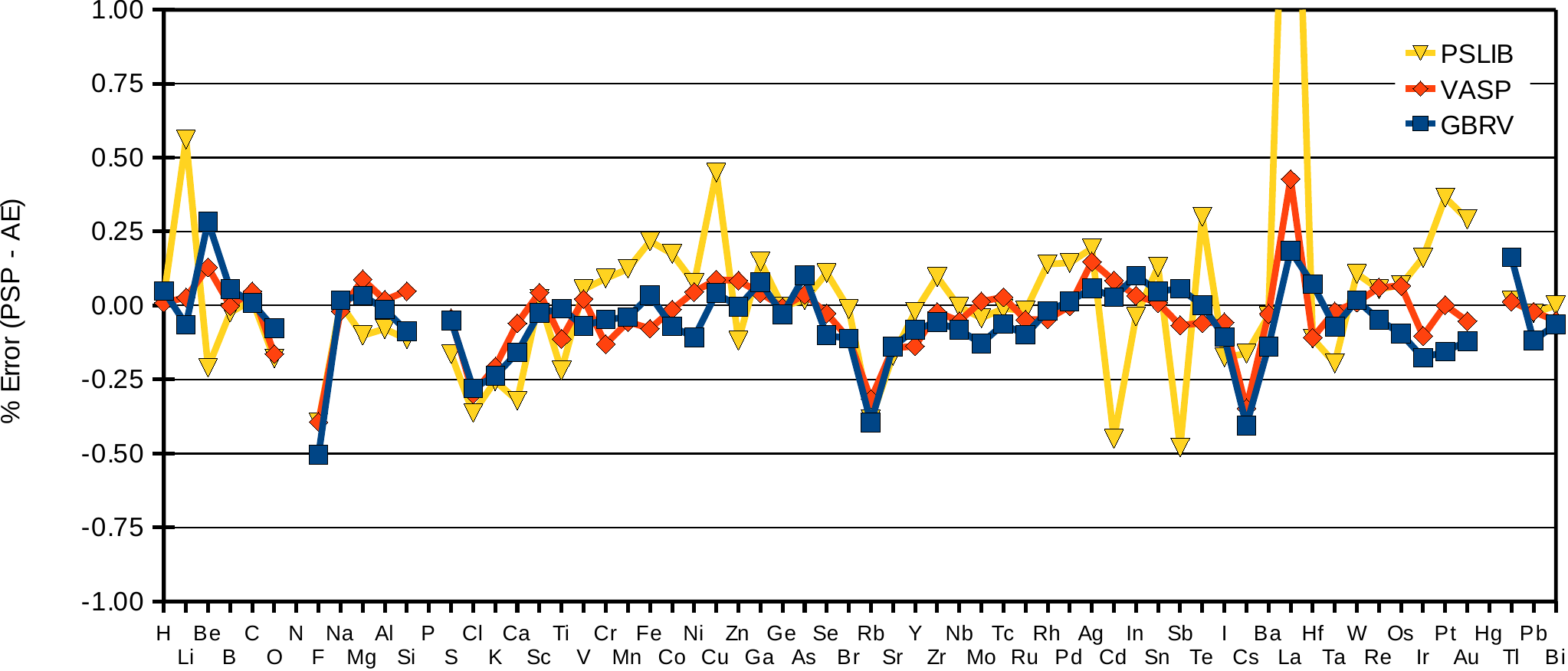}% Here is how to import EPS art
\caption{\label{fig:fcc} (Color online) Percent difference in AE versus
pseudopotential calculations for \textit{fcc} lattice constant.  GBRV results in blue squares, VASP in red diamonds, and
PSLIB potentials are yellow triangles.}
\end{figure*}

\begin{figure*}
\includegraphics[width=5in]{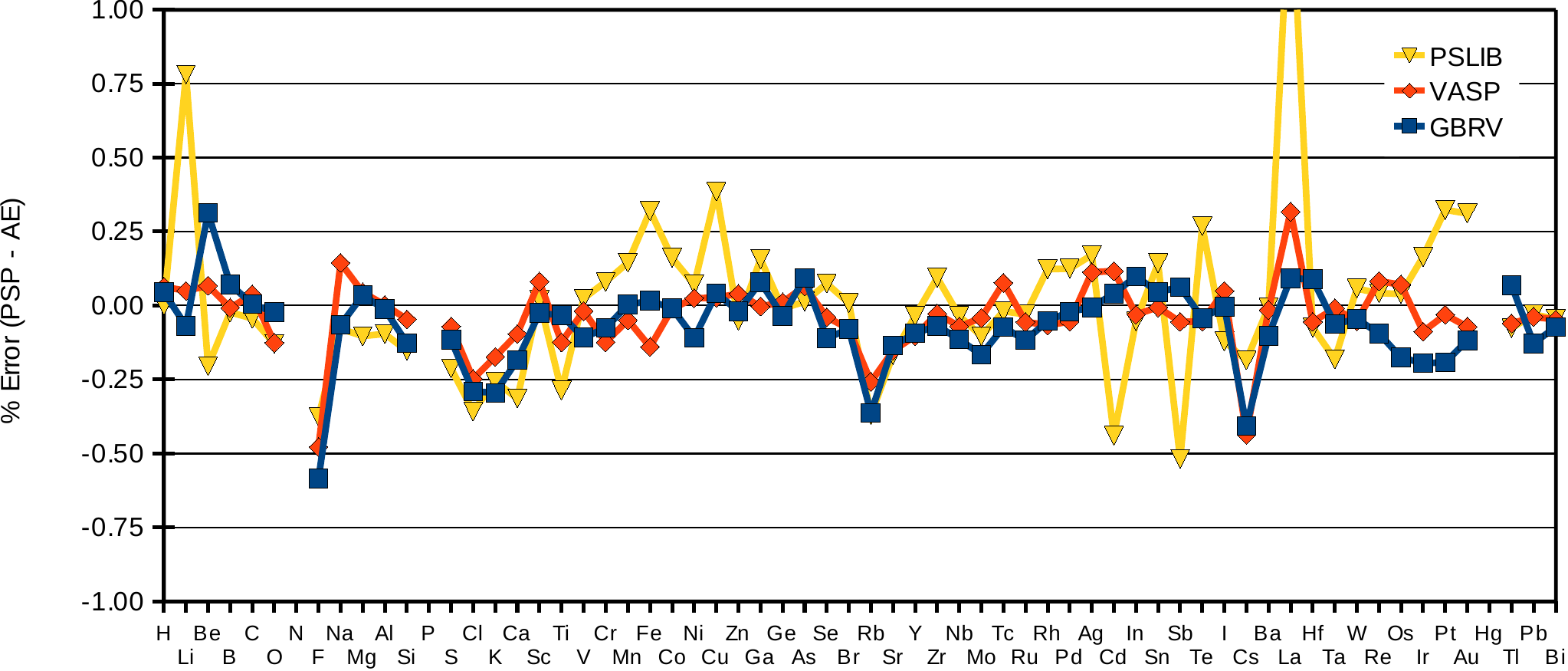}% Here is how to import EPS art
\caption{\label{fig:bcc} (Color online) Percent difference in AE versus
pseudopotential calculations for \textit{bcc} lattice constant.  GBRV results in blue squares, VASP in red diamonds, and
PSLIB potentials are yellow triangles.}
\end{figure*}

\begin{figure*}
\includegraphics[width=5.0in]{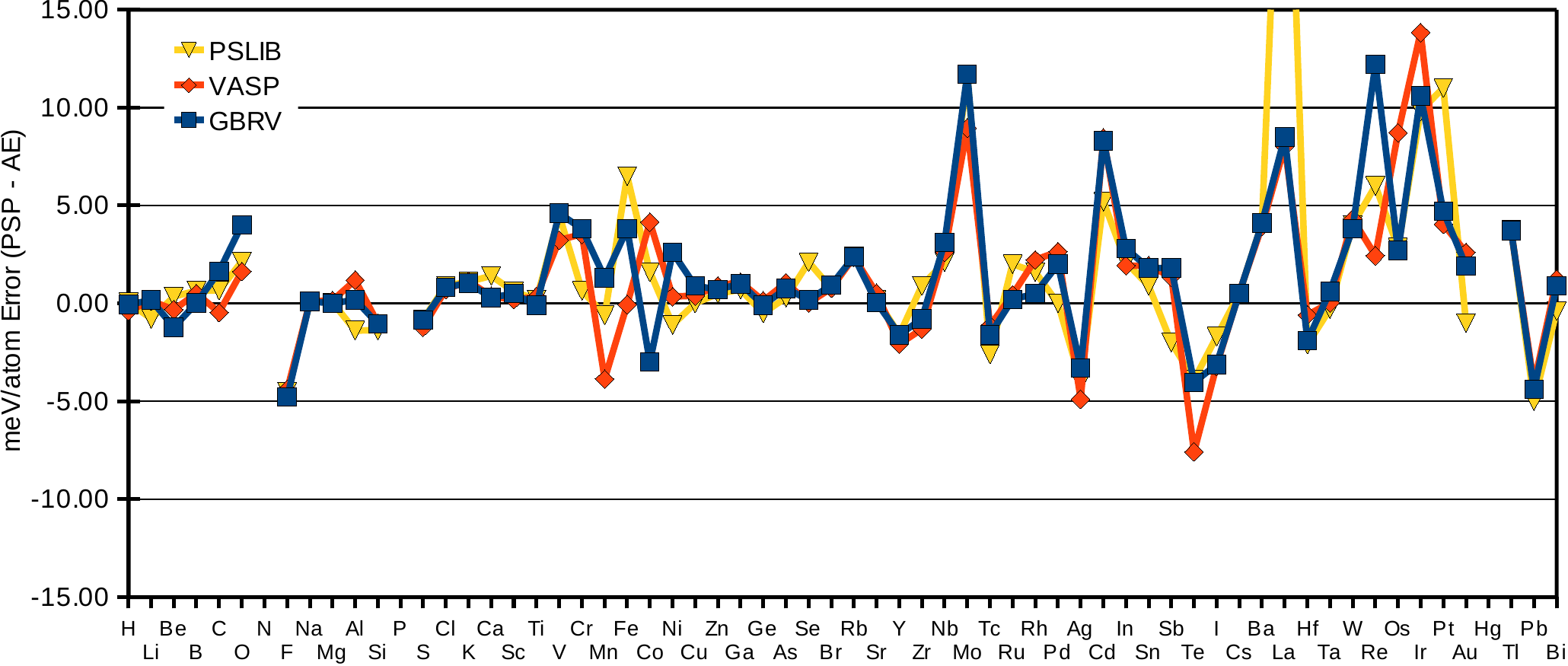}% Here is how to import EPS art
\caption{\label{fig:energy} (Color online) Error in energy difference (meV per
atom) between \textit{fcc} and \textit{bcc} structures.  GBRV are in blue squares, VASP in red diamonds,
and PSLIB in yellow triangles.  For context, the standard deviation of the AE energy differences is 0.21 eV.}
\end{figure*}

\begin{figure*}
\includegraphics[width=5.0in]{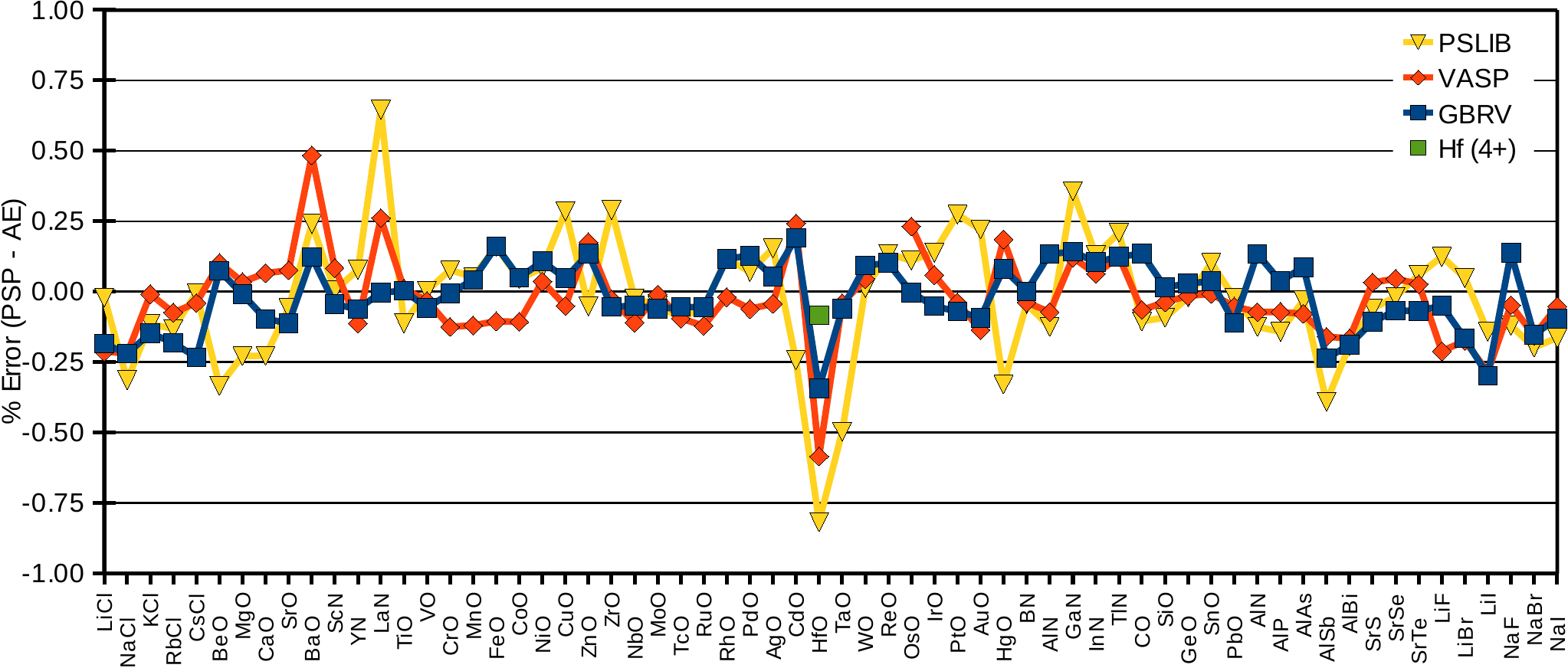}% Here is how to import EPS art
\caption{\label{fig:rocksalt} (Color online) Percent difference in AE versus
pseudopotential calculations for rocksalt lattice constants.  GBRV
potentials are blue squares, VASP potentials are red diamonds, PSLIB potentials are yellow triangles, and the Hf$^{4+}$ potential is an isolated green square (see text).  }
\end{figure*}

\begin{figure*}
\includegraphics[width=5.0in]{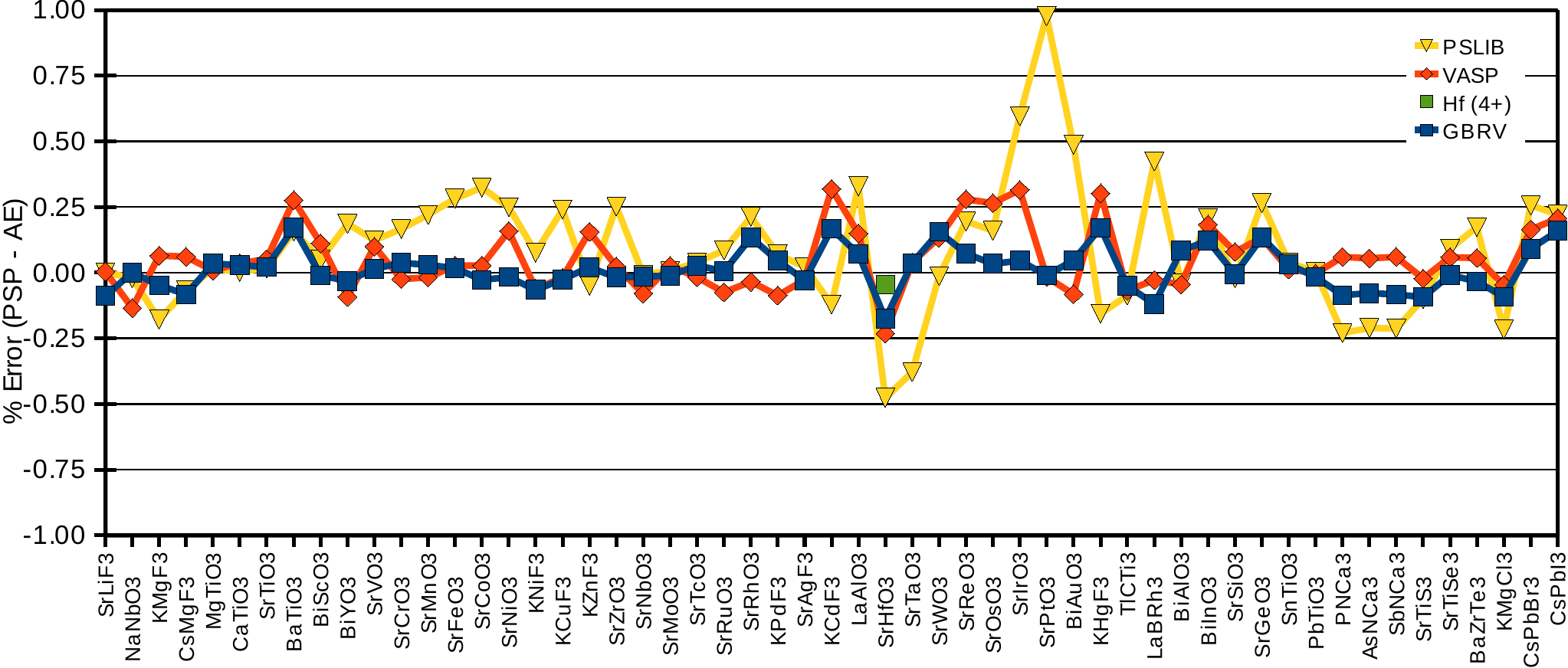}% Here is how to import EPS art
\caption{\label{fig:perovskite} (Color online) Percent difference in AE versus
pseudopotential calculations for perovskite (and anti-perovskite)
lattice constants.  GBRV potentials are blue squares, VASP potentials
are red diamonds, PSLIB potentials are yellow triangles, and the
Hf$^{4+}$ potential is an isolated green square (see text).}
\end{figure*}

Graphical representations for some of the tests reported in Table
\ref{tab:summary} are presented in
Figs.~\ref{fig:fcc}-\ref{fig:perovskite}.  Interestingly, the largest
average errors tend to be in the simplest compounds, especially the
\textit{fcc} and \textit{bcc} structures, and we note that the lattice
constant errors for these two structures are highly correlated with
each other (the correlation coefficient between the \textit{fcc} and
\textit{bcc} lattice constant errors for the GBRV potential set is
0.95).  In most cases the elements with large errors in
\textit{fcc} and \textit{bcc} lattice constants and bulk moduli tend
to have similar errors for all three pseudopotential sets, which
suggests that the errors are related to the frozen-core approximation
or some aspect of the all-electron calculations rather than any
specific problem with a specific potential.  The worst performing
potentials in this test tend to be from either the alkaline metals or
the halides, both of which tend to underestimate lattice constants.  F
and Cl, as well as several other first row elements including Li and
Be, can be improved by reducing the cutoff radii and accepting higher
plane-wave and charge density cutoffs; however, our current potentials
perform sufficiently well in most realistic compounds (see below).
The errors in the lattice constants of Rb and Cl may be related to the
frozen-core approximation, as the errors were consistent across all of the
pseudopotentials we constructed for these elements.  The errors in energy difference
between the \textit{fcc} and
\textit{bcc} structures, shown in Fig.~\ref{fig:energy}, tend to be
small and highly correlated across all three pseudopotential sets.  We note
that the largest errors tend to be for elements with a large energy separation
between the two structures, and that all of the calculations agree on the more
stable structure in all cases except for Pd, where the calculated AE energy difference is only $-$1.5 meV/atom.

\begin{figure}
\includegraphics[width=3.0in]{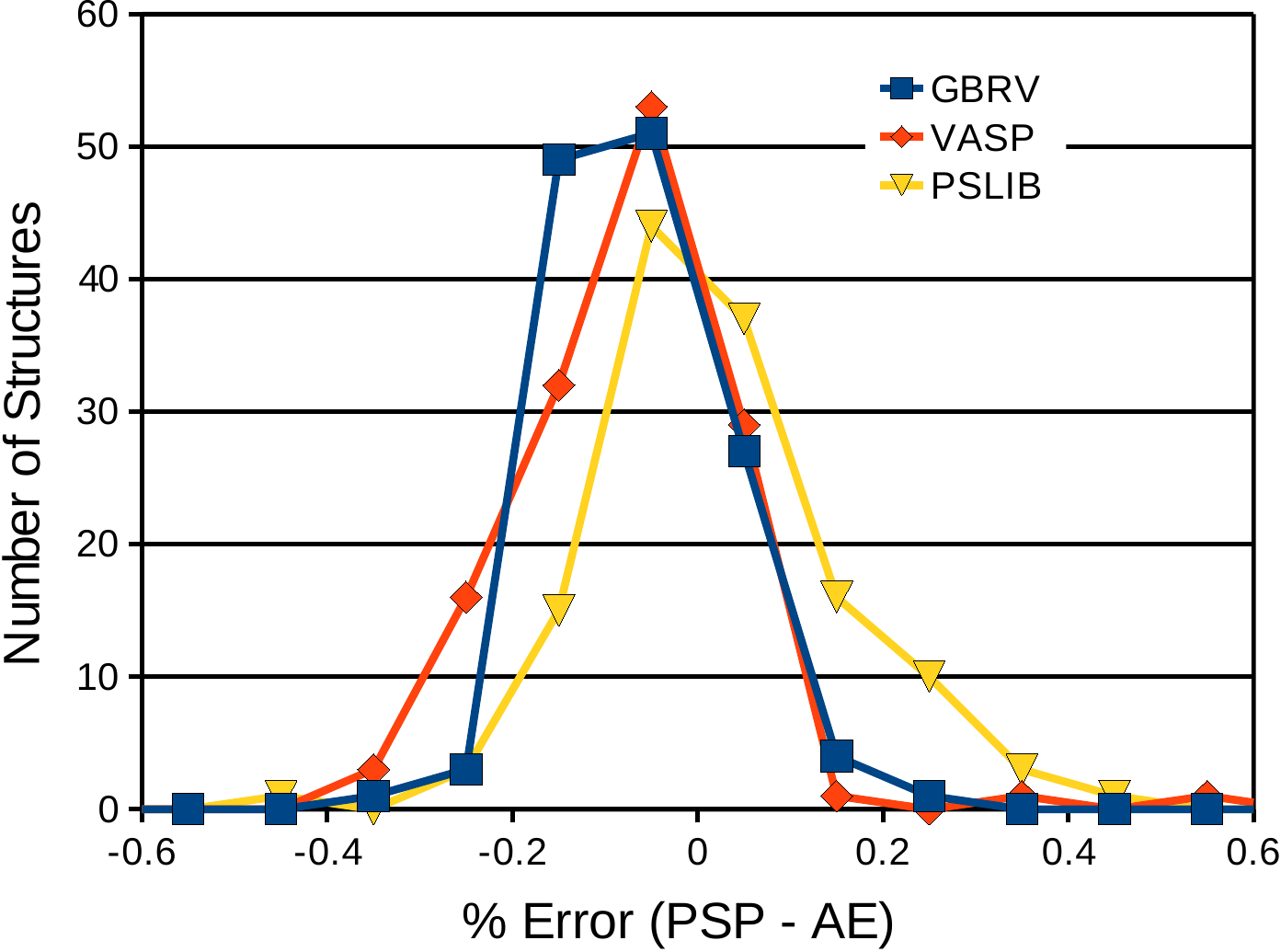}% Here is how to import EPS art
\caption{\label{fig:hh_hist} (Color online) Histogram of \% error in
lattice constants (PSP - AE) for the half-Heusler test set.  GBRV results in blue squares, VASP in red diamonds, and PSLIB in yellow triangles.}
\end{figure}

All of the potentials show good overall performance on most compounds in the rocksalt,
perovskite, and half-Heusler testing sets, as shown in
Figs.~\ref{fig:rocksalt}-\ref{fig:hh_hist} (see also supplementary
materials).  Apparently the ionic and covalent bonding of these
compounds is relatively insensitive to either the frozen-core
approximation or the details of pseudopotential construction, at least
for carefully tested potentials.  The GBRV potentials have the
advantage that they show fewer ``poor'' cases (defined as those
showing lattice constant errors exceeding $\pm 0.2 \%$), which makes them
particularly useful for a robust high-throughput study.

The GBRV potentials perform very well in the ionic NaCl and perovskite
structures, as shown in
Figs.~\ref{fig:rocksalt}--\ref{fig:perovskite}, with the most notable
exceptions being HfO and SrHfO$_3$.  Hf has a filled $4f$ orbital which
overlaps strongly both spatially and energetically with the valence
and semicore $s$, $p$, and $d$ orbitals, and this $4f$ orbital would
have to be included in any truly transferable Hf potential.
Unfortunately, including such a localized orbital is impossible within
our convergence criteria.  In order to treat these technologically
important oxides accurately, in the spirit of
Ref. \cite{sinisa}, we created a second Hf potential, generated
to reproduce a Hf$^{4+}$ ionic configuration.  This potential gives
improved performance in Hf oxides, as shown by the green square in
Figs.~\ref{fig:rocksalt}-\ref{fig:perovskite}; however, it gives
poor performance for metallic Hf (e.g., +0.71\% lattice constant error
in the \textit{fcc} structure) and the errors reported in table
\ref{tab:summary} all refer to the standard Hf potential.  The
remaining large errors in the ionic testing sets mostly involve
combinations of alkaline metals with halides (e.g., LiI or CsCl), which
we already noted are difficult to pseudize.

The performance of the GBRV potentials is also excellent for the large
half-Heusler data set, as shown in Fig. \ref{fig:hh_hist}.  The
lattice constant errors are approximately normally distributed, with a
small bias of $-$0.07\%, and only 5 out of 128 structures (SbTaCo,
MnSbOs, MnPdTe, LiAuS, and CdPLi) outside of the $\pm 0.2 \%$ lattice
constant error range, and many of those barely outside.  We note that
the compounds with large errors all contain large transition metals,
and that it is probably possible to improve Au, Cd, Pd, and Ta by
including additional semicore states if more accuracy is required.  
The ability to modify potentials is an important feature of open-source 
libraries such as GBRV.
Fig. \ref{fig:hh_hist} also shows that the VASP and PSLIB potentials
have tails of underestimated and overestimated lattice constants,
respectively, which contribute to their higher RMS errors.  Despite
these outliers, we note that the remaining lattice constant errors 
are highly correlated across potential sets, with a correlation
coefficient between the GBRV and VASP lattice constant errors of 0.63
(0.36 for GBRV and PSLIB).  This correlation, which can also be seen
in Figs. \ref{fig:fcc}--\ref{fig:perovskite} for the previous testing
sets, suggests that much of the remaining error is due to the
frozen-core approximation.  The errors in bulk modulus are even more
highly correlated, with a correlation coefficient for the half-Heusler
testing set of 0.92 between GBRV and VASP (0.84 between GBRV and
PSLIB).

As a reminder, all the results presented until now have been for
non-spin-polarized calculations.  In order to obtain some information
about the ability of the pseudopotentials to reproduce magnetic
properties, we have carried out calculations for the binary
transition-metal oxides in the rock salt structure
(Fig.~\ref{fig:rocksalt}), but now initialized in a ferromagnetic spin
configuration.  For those that converged to a ferromagnetic ground
state, we compare the resulting magnetic moments with the AE values in
Table \ref{tab:magmom}.  (All four calculations agreed as to which
materials remained ferromagnetic.)  The PSLIB potentials reproduce the
AE magnetic moments exceptionally well despite many elements having
fewer semicore states than the other potential sets, which we
attribute in part to their higher plane-wave and charge density
cutoffs.  The largest errors for the GBRV set are for NiO
and CoO.  We found that the magnetic properties of Ni in particular are very
sensitive to its inner cutoff radius.  Unfortunately, there is no
value for this parameter which both reproduces the magnetic moment of
NiO and is compatible with our convergence criteria; we had to
compromise and choose the best value compatible with our requirements.
We recommend significant testing before using any pseudopotential in
detailed magnetic calculations.

The strong correlation between the lattice constant errors in the
$fcc$ and $bcc$ structures, evident when comparing
Figs.~\ref{fig:fcc} and~\ref{fig:bcc}, suggests that it may be possible to
create an estimate of the lattice constant error of other structures
using this information.  In other words, perhaps each element is
consistently too small or too large across all structures, and this
error can be corrected for.  However, when we attempted to model the
lattice constant errors of the GBRV potential set as a linear
combination of the average \textit{fcc} and \textit{bcc} errors of the
elements in each compound, we found little improvement in RMS errors
beyond the improvement which came from subtracting the overall bias of
$-$0.06\%.  Similarly, a model of the lattice constant error based on a
global least squares fit to the testing sets, leaving out one compound
at a time to evaluate the fit, improved RMS errors less than 0.01\%.
In other words, the bias introduced by each pseudopotential depends
too strongly on chemical environment to be modeled with a single number.

\begin{table} 
\caption{\label{tab:magmom} Testing data for magnetic moments of
transition metal oxides with non-zero magnetic moments at the AE
non-spin polarized lattice constant.  All magnetic moments in
$\mu_{B}$ per primitive cell.}
%\begin{ruledtabular}
\begin{tabular}{lllll}
Compound   &  $\mu_{AE}$ & $\mu_{GBRV}$ & $\mu_{VASP}$ & $\mu_{PSLIB}$  \\
\hline
TiO   &   0.08  &  0.08 &   0.04 &   0.07 \\
VO    &   1.55  &  1.54 &   1.44 &   1.55 \\
CrO   &   2.91  &  2.98 &   2.97 &   2.90 \\
MnO   &   3.83  &  3.90 &   3.84 &   3.82 \\
FeO   &   3.83  &  3.85 &   3.83 &   3.83 \\
CoO   &   2.59  &  2.78 &   2.69 &   2.59 \\
NiO   &   1.82  &  1.75 &   1.01 &   1.83 \\
MoO   &   0.42  &  0.43 &   0.42 &   0.42 \\
TcO   &   2.02  &  2.01 &   2.01 &   2.02 \\
RuO   &   1.44  &  1.43 &   1.45 &   1.44 \\
RhO   &   0.36  &  0.36 &   0.37 &   0.36 \\
ReO   &   0.11  &  0.11 &   0.09 &   0.11 \\
OsO   &   1.79  &  1.79 &   1.79 &   1.79 \\
IrO   &   0.79  &  0.80 &   0.79 &   0.79 \\
\end{tabular} 
%\end{ruledtabular}
\end{table}

Finally, note that we have been treating the AE results from WIEN2k
as essentially exact, but this point deserves futher investigation,
as prelimiinary tests\cite{gian}
indicate that the differences between different AE codes can
sometimes be significant on the scale of our comparisons.

\subsection{GBRV PAW Library}

In order to achieve broader compatibility with open-source electronic
structure codes, specifically ABINIT, which can perform calculations
with PAWs but not with ultrasoft pseudopotentials, we use the
uspp2abinit\cite{uspp2abinit} add-on to the Vanderbilt ultrasoft code
to generate PAW versions of the GBRV library.  While closely related,
the formalism of PAW and ultrasoft potentials are not the
same\cite{paw2}, which results in differences between the two GBRV
potential sets which are larger than the differences between QUANTUM
ESPRESSO and ABINIT when run with identical norm-conserving
potentials.  For some elements which are particularly sensitive to
generation parameters, especially those with many semicore states or
core states which are close in energy to valence states, we had to
alter the generation parameters in order to make a PAW of equal
accuracy to the ultrasoft version\cite{abinit_notes}.  We find that the GBRV PAW
and ultrasoft libraries, tested with ABINIT and QUANTUM
ESPRESSO respectively, have the same overall accuracy relative to all-electron
calculations, and their errors in lattice constant are highly
correlated (e.g. the correlation coefficient between the perovskite
lattice constant errors is 0.73).

\section{Conclusions}

In conclusion, we have presented design criteria and testing results
for the new GBRV pseudopotential library optimized for high-throughput
DFT calculations.  We find that our potentials are both accurate and
transferable, performing well in tests of metallic, ionic, and
covalent bonding.  In addition, we have compiled testing results for
two PAW
libraries, which also perform well and demonstrate the reliability of
carefully designed pseudopotentials in electronic-structure
calculations.  While calculations using either ultrasoft
pseudopotentials or PAWs are more complicated to implement than those
using norm-conserving potentials, almost all modern electronic-structure
codes are now capable of using such potentials, with current code
development efforts continually expanding the set of features that
are compatible with them.  In particular, the GBRV potentials, available at http://physics.rutgers.edu/gbrv, 
can be used directly with the actively-developed open-source QUANTUM ESPRESSO and ABINIT packages.

For the properties investigated in this work, the GBRV potential set
provides better accuracy and robustness than the VASP or
PSLIB PAW sets, and at lower computational cost.  
In particular, the
GBRV potentials are designed to run at a plane-wave cutoff of 40 Ry
and a charge-density cutoff of 200 Ry, which are at least 25\% lower
than many of the PSLIB PAWs as well as lower than three VASP PAWs when
using the `high' precision setting recommended for variable cell
relaxations.

The GBRV potentials (like the PSLIB
potentials) also have the advantage of being open-source, which
allows calculations to be easily replicated and enables the potentials
to be modified and improved as needed by the electronic structure
community.  Furthermore, open-source potentials can be used with
open-source electronic structure packages, which have active
development communities and, like the potentials themselves, can
be improved and expanded upon as necessary.
% to complete future projects.

We hope that both the GBRV potential library itself, as well
as the design criteria and testing methodology presented here, will
improve the use and reliability of pseudopotential-based
high-throughput DFT calculations for a variety of materials design
applications.

%\begin{acknowledgments}
This work work was supported by NSF Grant DMR-10-05838
and ONR Grants N00014-09-1-0302 and N00014-12-1-1040.
We wish to thank D.R.~Hamann for valuable discussions.
%\end{acknowledgments}

%\bibliography{psp}% Produces the bibliography via BibTeX.

\end{document}